# Laser Driven Ultra-compact Undulator for Synchrotron Radiation

Junhao Tan[1,2], Yifei Li[1,2], Baojun Zhu[1,2], Changqing Zhu[1,2], Jinguang Wang[1,2], Dazhang Li[1,2], Liming Chen[1,2,3,5,*]

[1] *Beijing National Laboratory of Condensed Matter Physics, Institute of Physics, Chinese Academy of Sciences, Beijing 100190, China*

[2] *School of Physical Sciences, University of Chinese Academy of Science, Beijing 100190, China*

[3] *IFSA Collaborative Innovation Center, Shanghai Jiao Tong University, Shanghai 200240, China*

[4] *Institute of High Energy Physics, Chinese Academy of Sciences, Beijing 100049, China*

[5] *Songshan Lake Materials Laboratory, Dongguan, Guangdong 523808, China*

*\*lmchen@iphy.ac.cn*

**Abstract**

Laser wakefield accelerators have emerged as a promising candidate for compact synchrotron radiation and even x-ray free electron lasers. Today, to make the electrons emit electromagnetic radiation, the trajectories of laser wakefield accelerated electrons are deflected by transverse wakefield, counter-propagating laser field or external permanent magnet insertion device. Here, we propose a novel type of undulator which has a few hundred microns of period and tens of Tesla of magnetic field. The undulator consists of a bifilar capacitor-coil target which sustains strong discharge current that generates helical magnetic field around the coil axis when irradiated by a high energy laser. Coupling this undulator with state-of-the-art laser wakefield accelerators can, simultaneously, produce ultra-bright quasi-monochromatic x-rays with tunable energy ranging 5-250 keV and optimize the free electron laser parameter and gain length compared with permanent magnet based undulator. This concept may pave a path toward ultra-compact synchrotron radiation and even x-ray free electron lasers.

**Introduction**

Since proposed by T. Tajima and J. M. Dawson[1] in 1979, laser wakefield accelerators (LWFA) have made tremendous progress in electron acceleration and x-ray light sources. Electrons can be trapped and accelerated in the wakefield driven by an intense femtosecond (fs) laser with typical accelerating electric field of 100 GV/m[2, 3], possessing the inherent fs time duration and tens of pC charge, thus reaching a few kA peak beam current. Electrons with GeV energy[4, 5], ultra-high brightness[6], and kA beam current[7, 8] have been obtained experimentally.

LWFA can also be used to generate betatron and undulator radiation. Despite the longitudinal acceleration, electrons can simultaneously undergo transverse betatron oscillations in the wakefield which lead to the so called betatron radiation. Betatron x-ray pulses usually have broadband spectrum[9-11] since the transverse wakefield is so strong that acts as a wiggler. Undulator radiation combing permanent magnet and LWFA has also been demonstrated for visible light[12] and soft x-ray[13]. In such experiments, electrons accelerated by LWFA were injected into a permanent magnet undulator placing tens of centimeters downstream the gas jet.

For this type of undulator, the typical values for period and magnetic field are of 1cm and 1 Tesla, respectively[14]. Further, owing to the large divergence of LWFA electron beams (~ mrad), the electron flux would be reduced by few orders of magnitude after propagating through the long gap between the accelerator and undulator[15]. Thus, the x-ray brightness is strongly limited, e.g., $1.3 \times 10^{17}$ photons/s/mrad$^2$/mm$^2$/0.1%BW as reported by M.Fuchs *et.al*[13].

Shaped quasi-static strong magnetic field can be generated by laser driven capacitor-coil target[16-18], which is comprised of two parallel metal plates connected with a shaped coil, as schematically shown in Fig. 1. The two plates form a capacitor and drive a quasi-static (time scale of a few ns) current through the coil when one is irradiated by a high-energy nanosecond (ns) laser propagating through a hole on another plate, which collects the hot electrons emitted from the laser-produced plasma. Magnetic field exceeding hundreds of Tesla and lasting tens of ns is typical for such experiments[17, 19, 20].

In this paper, we present an elliptical undulator based on a bifilar shaped capacitor-coil. A bifilar coil[21] containing DNA-like double helical windings presents a helical magnetic distribution on axis when currents with equivalent amplitude flowing in opposite directions in those two wingdings. When synchronized LWFA electron beam is generated upstream the coil, the strong quasi-static magnetic field acts as an elliptical undulator and will bent the trajectories of electrons injecting along the coil axis thus produce synchrotron radiation. Significantly, the period of this laser driven elliptical undulator can be as short as a few hundred microns and the amplitude of the magnetic field as high as tens of Tesla, while the gap between LWFA and the undulator can be as short as a few millimeters. Combing this short period high strength undulator with LWFA will make better use of the high electron flux of LWFA and lead to a much higher x-ray brightness.

**Results**

*Principle of the short period high strength undulator*

The proposed scheme of the undulator and coupled x-ray source is presented in Fig. 1. We consider a laser with 1 ns pulse duration, 1 kJ energy and 1.06 µm wavelength is focused onto one of the copper plates (anode) with a focal spot size of 50 µm. Another copper plate (cathode), which collects the expanding hot electrons, is separated 600 µm from the first one, and the diameter of the two plates is 3.6 mm, similar to that used in current experiments[17, 20]. One can expect discharge currents with opposite directions in the two helices since the charge separation will build up an electrical potential between the two plates and the capacitor acts as a voltage source[22]. The internal diameter, helix pitch, wire diameter, and number of turns of the copper bifilar coil are 500 µm, 500 µm, 50 µm, and 20, respectively. The LWFA is coaxially placed upstream the coil with 1 mm gap, and the initial point of the electron beam is defined as origin (x = 0, y = 0, z = 0) of the whole system. The helix pitch defines the undulator period $\lambda_u = 500$ µm while the undulator strength is determined by the magnetic field amplitude on axis.

The temporal evolution of the coil current can be obtained since the capacitor coil can be described as a RLC electric circuit which is governed by the laser and target geometry[23, 24]. For the capacitor coil considered above, the resistance, self-inductance, mutual inductance and the capacitance are $R = 0.158$ Ω, $L = 6.0$ nH, $M_{12} = M_{21} = -0.5$ nH and $C = 0.184$ pF, respectively. It is known that plasma dynamics in the capacitor is essential but difficult for the coil current modeling, and we estimate the coil current according to the model developed by T.K.Tikhonchuk et al[24]. This model, accounting for the space charge neutralization and plasma magnetization between the capacitor plates, shows good validity when compared with former experimental data. For our case, this model indicates a current with 36.0 kA peak amplitude, $\leq 1$ ns raising time and $\tau_{re} = L/R \sim 38.0$ ns relaxation time can be generated. In principle, we can realize the variation of undulator strength by changing the time delay between the fs laser and ns laser.

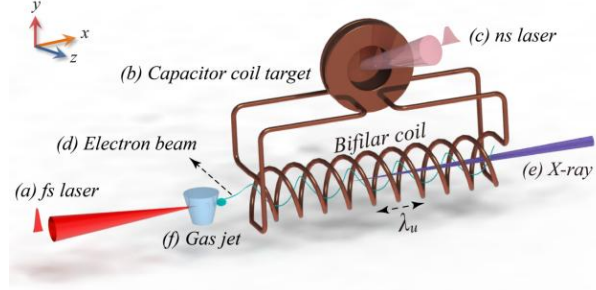

Fig.1. Schematic diagram of the bifilar elliptical undulator and the coupled x-ray source. (a) Intense femtosecond laser is focused in a gas jet to drive a wakefield and accelerate electrons. (b) Capacitor-Coil target formed by a copper bifilar coil and two copper plates. (c) High energy nanosecond laser is focused on one plate through a hole on another. (d) Electrons accelerated in the wakefield present elliptical orbits after injecting into the undulator along the coil axis. (e) Ultra-bright x-ray will be emitted. The coil pitch defines the undulator period $\lambda_u$ while the undulator strength is determined by the magnetic field amplitude on axis. The gap between the LWFA and the capacitor in x can be as short as one millimeter thus make better use of the high flux of LWFA electron beams.

The spatial field pattern is essential for a magnetic undulator. For an ideal bifilar coil, near the coil axis (y = 0, z = 0), the axial magnetic field $B_x$ vanishes while the transverse component $B_y, B_z$ are sinusoidal with the period equals to the coil pitch $\lambda_u$. The static 3D field distribution of the bifilar coil is calculated numerically and the whole coil geometry, including the connection wires between the coil and copper plates, has been taken into account with the undulator length $L_u = 12.0$ mm. The calculation is based on the fact that the time interval ($\tau_e \sim L_u/c \sim 40.0$ ps) of relativistic electrons propagating through the whole undulator is negligible small compared with relaxation time $\tau_{re}$ of the coil current (38.0 ns), thus the current amplitude can be regarded as a constant and static magnetic field distribution is assumed. The field map in Fig. 2 shows the magnetic field in the horizontal plane (y = 0) under a 30 kA coil current. Figure 2 (a), (b), (c) represents $B_x, B_y, B_z$, respectively. The map apparently shows an alternating polarity magnetic field distribution with 500 μm period determined by the coil pitch. And, along the axis, magnetic field presents an elliptical distribution, i.e., $B_x$ equals to zero while $B_y, B_z$ are sinusoidal[14]. The magnetic field experienced by one typical electron distinctly confirms the elliptical field pattern, as illustrated by the purple dotted line in Fig. 2. It should be noted that taking the whole coil geometry into account is vital for accurately modeling the undulator field. For a symmetric bifilar coil, the field amplitude $B_{y0}, B_{z0}$ for y, z directions should be exactly equal with each other. For our case, however, coil asymmetry originating from connection wires results in the average field amplitude $B_{y0}$ = 16.3 T and $B_{z0}$ = 14.8 T, respectively. Further, the magnetic field decreases to zero at 1 mm away from the coil, thus the gap between LWFA and undulator can be set to as short as 1 mm without net effect on each other. This will allow making full use of high electron flux of LWFA and eventually lead to an increase of x-ray brightness[15].

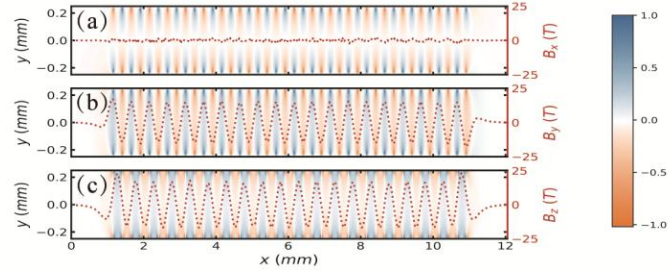

Fig.2. Magnetic field distribution with a coil current of 30 kA. (a), (b), (c) represents $B_x, B_y, B_z$ with the same scale, respectively, and the coordinate is shown in Fig. 1. The field map shows alternating-polarity magnetic field distribution with the period equivalent with coil pitch. The purple dotted line represents the magnetic field experienced by a typical electron and shows an elliptical field pattern. The axial part $B_x$ equals to zero while $B_y, B_z$ are sinusoidal. The averaged amplitude of $B_y$ and $B_z$ are 16.3 T, 14.8 T, respectively.

Electron dynamics inside the undulator is also critical for the operation of the coupled x-ray source. Figure 3 shows transverse motions of a typical electron inside the undulator generated with a 30 kA coil current. LWFA electrons are injected into the undulator along the coil axis. Electron presents an elliptical orbit with a period of 500 μm, as shown in Fig. 2 (b), (c). It should be mentioned that the sinusoidal electron motion is combined with small transverse deflection and drift as the result of field inhomogeneity, i.e. electron position in z direction drifts 0.57 μm at the exit (x = 12 mm) of the undulator as illustrated in Figure 2 (a). This drift is consistent with the value 0.62 μm given by the second field integral[14] $z = \frac{e}{\gamma m_0 c} II_y = \frac{e}{\gamma m_0 c} \int_{-\infty}^{+\infty} \int_{-\infty}^{s} B_y(s')ds'ds$, where $e, m_0, \gamma$ and $c$ are the elementary charge, the electron rest mass, the Lorentz factor and the speed of light in vacuum, respectively. For lower electron energy, this drift will increase but still negligible ($< 5$ μm) when compared with the coil diameter for electrons considered in this paper. Electron beams will maintain this small drift and defocus due to the initial divergence when moving in the undulator. The corresponding electron beam phase space distribution at the undulator entrance and exit are shown in Fig. 4. The ellipse at the undulator exit, Fig. 4(b), (d), extending from the left lower quadrant to right upper quadrant indicates a divergent electron beam. And after propagating through the undulator, the beam radius (root mean square, rms) in y and z direction grow up from 3.0 μm to 5.19 μm and 5.30 μm, respectively, while the divergence is consistent with the initial one. This growth in beam radius is acceptable for the purpose of undulator radiation, while, however, toward x-ray light sources where extensive undulator periods are required, i.e. x-ray free electron lasers (X-FEL), magnet devices for carefully manipulating electron phase space distribution are needed.

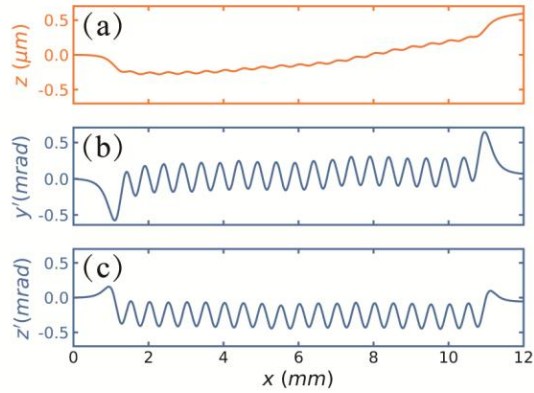

Fig.3. Transverse motions of a typical electron. (a), (b), (c) represent z, $y' = \frac{p_y}{p_x}$, $z' = \frac{p_z}{p_x}$ respectively, where z is the electron position in one transverse direction and $p_i$ is the electron momentum in corresponding direction. (a) Electron undergoes a transverse drift caused by field inhomogeneity, i.e., z drifts $0.57 \mu m$ at the exit of the undulator, which agrees well with the value predicted by the second field integral. (b), (c) Electron transverse momentum normalized to longitudinal momentum. Apparently, electron exhibits an elliptical motion inside the undulator. The electron parameters are consistent with the UTEXAS experiment result listed in Table.1.

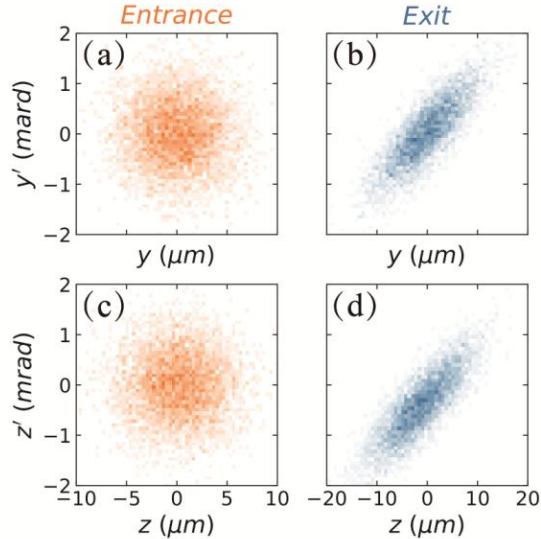

Fig.4. Electron beam transverse phase space distribution. (a), (c) Electrons are Gaussian distributed at the undulator entrance (x = 0) with the rms beam radius of $3 \mu m$ and the rms divergence of 0.6 mrad. (b), (d) After propogation through the undulator, electrons present a defocusing elliptical distribution with the rms beam radius and divergence of $5.3 \mu m$ and 0.6 mrad, respectively. For this calculation, $5 \times 10^3$ electrons were used.

## X-ray radiation coupling the undulator with LWFA

To get further insight of the features of the x-ray source, we modelled the coupled system by synchrotron radiation code WAVE (see **Materials and methods**). Insertion devices can be characterized in terms of period and magnetic field amplitude, and can be summarized by the strength parameter, $K = eB_0\lambda_u/2\pi m_e c$, where $B_0 = \sqrt{B_{x0}^2 + B_{y0}^2}$ is the on axis magnetic field amplitude for elliptical undulator and $\lambda_u$ is undulator period as mentioned above. For the proposed undulator with $\lambda_u$ = 500 μm, $B_0$ = 22.02 T, $K$ equals to 1.07. The strength paramater differs undulator ($K \leq 1$) and wiggler ($K \gg 1$) regime in terms of photon energy and yield[25]. In the undulator regime ($K \leq 1$), electrons radiate around the fundamental wavelength $\lambda_\gamma = \lambda_u(1 + K^2/2)/2\gamma^2$ with the number of photons $N_\gamma = 2\pi\alpha K^2/3$ emmited per magnet period, where $\lambda_\gamma$ is on axis fundamental radiation wavelength, $\alpha \approx 1/137$ is the fine structure constant and $K$ is the strength parameter. At a fixed electron energy $\gamma$, shorter radiation wavelength can be obtained with short undulator period $\lambda_u$ while high undulator strength will enlarge the photon yield. To verify the validity of the proposed x-ray source, undulator radiation coupling this undulator and state-of-the-art LWFA is calculated. We take electron beam parameters close to that obtained experimentally by three groups (W.P. Leemans et al. at LBNL[4], X. Wang et al. at UTEXAS[5] and W. Wang et al. at SIOM[6]) into account and summarize electron pamameters in Table Ⅰ. In the case of LBNL electron beam, x-ray pulse peaked at 252.7 keV with 59.7% (FWHM) relative energy spread (RES) and a peak brightness of $3.13 \times 10^{25}$ photons/s/mrad$^2$/mm$^2$/0.1%BW can be generated. This brightness is comparable and even higher than that of the third generation sychrotron radiation based on energy recovery linear accelerators but at a much higher photon energy. The Stokes vector at peak photon energy is (1, -0.07, 0.007, 1), which suggest the x-ray is almost pure right-hand circularly polarizd[14]. This relatively large x-ray RES mainly comes from the large RES (6%) and divergence of the electron beam since they contribute to the spectrum broadening as[26] $(\Delta\lambda_\gamma/\lambda_\gamma)^2 = (2\Delta\gamma/\gamma)^2 + (\gamma^2\epsilon^2/\sigma^2)^2 + (1/N)^2$, where $\epsilon$ and $\sigma$ are the emmitance and transverse beam radius of the electron beam. And x-rays with RES down to 15.1% (FWHM) can also be generated by electron beam of SIOM case thanks to the small electron RES (< 1%).

**Table 1 Undulator and electron beam parameters**

| *Undulator* | | | | |
|---|---|---|---|---|
| Period | Coil current | | Strength parameter $K$ | |
| 500 μm | 20 kA[3] | | 0.71 | |
| *Electron beam parameters for spectrum calculation*[d] | | | | |
| | Average energy | Energy spread | Divergence | Charge |
| SIOM | 0.6 GeV | 0.9%[a] | 0.3 mrad[a] | 42 pC |
| UTEXAS | 2.0 GeV | 6%[b] | 0.6 mrad[b] | 63 pC |
| LBNL | 4.2 GeV | 6%[c] | 0.3 mrad[a] | 6 pC |

[a]rms values. [b]FWHM values. [c]The coil currrent is chosen for higher x-ray photon energy. [d]The duration and size of electron beams are absent in these experimental papers and are assumed to be 10 fs and 3 μm (at x = 0), respectively.

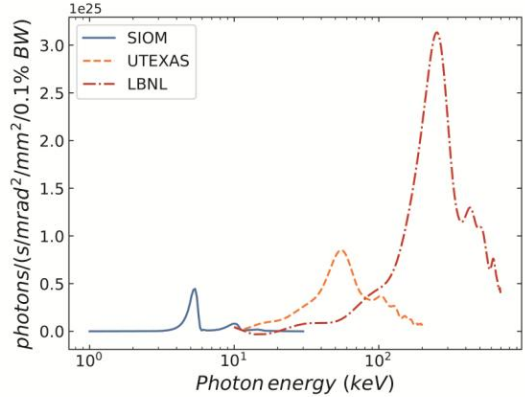

Fig.5. Undulator radiation spectrum by state-of-the-art LWFA. Electron beams are injected into the undulator at $x = 0$ with the initial parameters sampling from Gaussian distribution. The yellow solid, red dashed and blue dot-dashed line denote electron beams with parameters close to that obtained by SIOM, UTEXAS, and LBNL, respectively.

On the other hand, tunabilities in photon energy and brightness are of key importance for applications of x-ray light source. In synchrotron radiation science, a common way of tuning the photon energy is variation of the magnetic field amplitude $B_0$ since the variation of electron energy is rather complicated and even impossible due to the focusing and deflecting devices. For the undulator proposed, variation of $B_0$ can be obtained by tuning the coil current. This can be realized either by varying the time delay between the fs and ns laser or by variation of the energy of the ns laser[27]. Magnetic field amplitude $B_0$ under various coil current and corresponding $K$ are shown in Fig. 6 (a). The field amplitude increses linearly with the coil current, and as a rule of thumb 1.0 kA equals to 0.74 T for our capacitor coil geometry. The strength parameter $K$ can be varied from 0.34 to 1.07 while the coil current increses from 10 kA to 30 kA. And the corresponding peak photon energy can be varied from 17.3 keV to 12.2 keV when we take 1 GeV electrons as an example, as illustrated in Fig. 6 (b). The variation of electron energy presents a more efficient way since the coupled x-ray source is free of other beam manipulating devices and the dependence of photon energy and total power radiated on electron energy scales as $\gamma^2$. Peak x-ray photon energy increse from 4.2 keV to 59.2 keV while the electron energy ranging from 0.6 GeV to 2.2 GeV with 30 kA coil current, blue dot-dashed line in Fig. 6 (b). Higher photon energy can also be obtained with lower coil current on the cost of lower x-ray brightness, i.e., 6.3 keV to 84.9 keV under 10 kA coil current with the same electron energy range above, red line in Fig. 6.

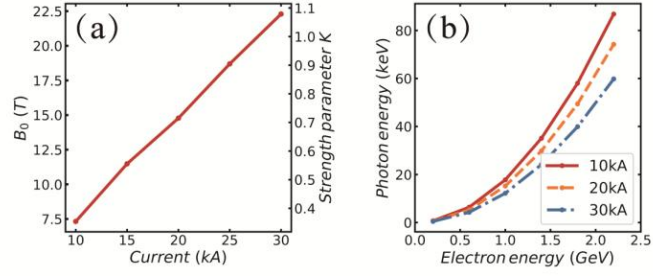

Fig.6. Tunability of the coupled x-ray source. (a) Amplitude for $B_0$ under different coil current. The strength of the undulator can be tuned by variation of the coil current, e.g. by varying the time delay between the fs and ns laser. (b) Peak energy of x-ray photon emitted by electrons with different energy or coil current.

**Disscussion**

So far we have shown the validity of the undulator and performance of the coupled x-ray sources. Owing to its intrinsic ultra-high acceleration gradient, kA electron beam current and fs time duration, LWFA is regarded as a promising candidate for compact X-FEL. For the FEL instability to occur, conditions applied on the electron beams can be characterized by the FEL pierce parameter $\rho = \frac{1}{2\gamma}(I/I_A)^{1/3}(A_u\lambda_u/2\pi\sigma)^{2/3}$, where $I$ is the electron beam current, $I_A = 4\pi\epsilon_0 mc^3/e \approx 17.0$ kA is the Alfvén current, $\sigma$ is the rms transverse electron beam radius and $A_u = K[J_0(x) - J_1(x)]/\sqrt{2}$ with $x = K^2/4(1 + K^2/2)$, $J_0$ and $J_1$ are Bessel functions. The RES of electron beam has to be smaller than $\rho$, of which for permanent magnet undulator based X-FEL is typically below $10^{-3}$, and, distinctly, this is rather strict for even the state-of-art LWFA with RES around 1%. Thus one may conclude that toward a compact X-FEL driven by LWFA, $\rho$ has to be as high as possible, while the 1D power gain length $L_{g_{1D}} = \lambda_u/4\pi\sqrt{3}\rho$, defined by $P \propto exp(x/L_{g_{1D}})$, should be minimized, where $P$ is the radiation power.

It is interesting to compare the $\rho$ and $L_{g_{1D}}$ between the coupled system and conventional radiofrequency accelerators based X-FEL. The dependence of $\rho$ and $L_{g_{1D}}$ on the field strength $B_0$ is illustrated in Fig. 7. For this comparison, we consider electron parameters for both types of accelerators on the prerequisite of fixed radiation photon energy 1 keV. The red lines denote $\rho$ and $L_{g_{1D}}$ for the coupled system while the blue dashed lines represent parameters for the radiofrequency accelerator based hard x-ray FEL SACLA. Generally, one can observe that $\rho$ increases with the magnetic field amplitude $B_0$ while $L_{g_{1D}}$ decrease. And $L_{g_{1D}}$ for LWFA (a few cm) is three orders of magnitude shorter than that of radiofrequency accelerator (a few m), which confirms the idea that LWFA driven FEL can be rather compact. Further, thanks to the short period and high field strength, the undulator we proposed is beneficial for both larger $\rho$ and shorter $L_{g_{1D}}$. That is, to generate x-rays of a fixed energy, short period will lower the electron energy required and shorten $L_{g_{1D}}$ owing to the linear dependence of $\lambda_\gamma$ and $L_{g_{1D}}$ on $\lambda_u$, while the lower electron energy and high magnetic field strength increasing $\rho$. The $\rho$ and $L_{g_{1D}}$ for the proposed system, red stars in Fig. 7, are $\rho = 4.12 \times 10^{-3}$ and $L_{g_{1D}} = 5.49$ mm, respectively. This value of FEL pierce parameter $\rho$ is close to electron beam RES of state-of-the-art LWFA while the gain length is only half of the undulator length. For self-amplified spontaneous emission (SASE) FEL, the radiation power typically saturates at about $20L_{g_{1D}}$ when taking 3D effects such as diffraction and emittance into account[26]. For the undulator proposed, $20L_{g_{1D}}$ is only 10.98 cm, corresponding to $N_s = 11$ segments of the undulator. This simple estimation shows the potential advantages of applying this undulator to compact X-FEL driven by LWFA.

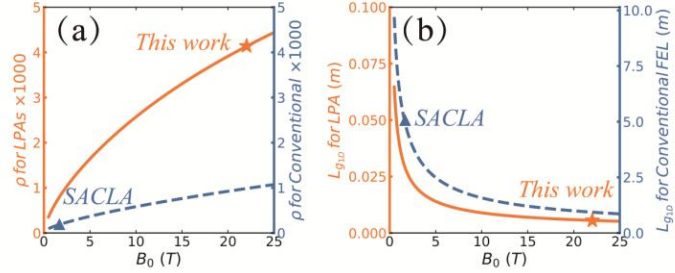

Fig.7. (a) FEL pierce parameter and (b) 1D FEL gain length for LWFA and conventional accelerators. The red solid and blue dashed line represent $\rho$ and $L_{g1D}$ for LWFA and conventional accelerators (SACLA), respectively, under different undulator strength. On the prerequisite of emitting 1 keV x-ray photons, The blue triangles correspond to the undulator used in SACLA ($1.68T$) while the red stars denote the proposed undulator combined with LWFA. The energy, initial beam radius, duration and charge of the LWFA electron beam are 300 MeV, 3 μm, 10 fs and 60 pC, respectively.

In summary, we have presented a short period high strength elliptical undulator using a bifilar shaped capacitor coil target. With currently available laser parameters, we show that an elliptical undulator with 500 μm period and up to 22.0 T magnetic field amplitude can be generated. Undulator radiation by coupling this undulator with laser plasma accelerators was also presented. We demonstrate that right-hand circular polarized x-ray with tunable energy ranging $5 - 250$ keV and high brightness of the order of $10^{25}$photons/s/mrad$^2$/mm$^2$/0.1%BW can be produced. For this coupled system, FEL pierce parameter can be increased close to the RES of state-of-the-art LWFA electron beam while the gain length is short down to a few millimeters. This concept can be further improved, for example, by using micro-fabricated capacitor coil with tens of microns period thus requiring less energetic nanosecond laser. This may lead to an ultra-compact synchrotron radiation and even X-FEL, not only the accelerator and undulator parts but the footprint of the whole system.

**Materials and methods**

*Magnetic field pattern of bifilar coil*

For an ideal bifilar coil, the magnetic field near the coil axis can be expressed analytically as[21]:

$$B_y = -B_0 \left\{ \left[1 + \frac{1}{8}k^2(3y^2 + z^2)\right] sin(kx) - \left(\frac{1}{4}k^2 yz\right) cos(kx) \right\} \quad (1)$$

$$B_z = B_0 \left\{ \left[1 + \frac{1}{8}k^2(y^2 + 3z^2)\right] cos(kx) - \left(\frac{1}{4}k^2 xy\right) sin(kx) \right\} \quad (2)$$

$$B_x = -B_0 \left[1 + \frac{1}{8}k^2(y^2 + z^2)\right][y\cos(kx) + z\sin(kx)] \quad (3)$$

where $k = 2\pi/\lambda_u$ is the undulator wave number, $B_0$ is the transverse field amplitude on axis and is associated with the coil current. One can immediately observe that, near the coil axis (y = 0, z = 0), the axial magnetic field $B_x$ equals to zero while the transverse parts $B_y$ and $B_z$ are sinusoidal with the period equivalent to the coil pitch $\lambda_u$. The 3D magnetic field distribution was calculated numerically using electromagnetic simulation software in the magnetostatic mode.

*Numerically modelling of the coupled system*

The LWFA electrons are injected into the undulator with the initial parameters sampling from Gaussian distribution as:

$$f(y, y', z, z', x, E) = \frac{1}{(2\pi)^3 \sigma_E \sigma_x \epsilon_y \epsilon_z} \exp\left[-\frac{\gamma_y y^2 + 2\alpha_y y y' + \beta_y y'^2}{2\epsilon_y} - \frac{\gamma_z z^2 + 2\alpha_z z z' + \beta_z z'^2}{2\epsilon_z} - \frac{(E-E_0)^2}{2\sigma_E^2} - \frac{x^2}{2\sigma_x^2}\right] \quad (4)$$

where $\alpha_{y,z}, \beta_{y,z}$ and $\gamma_{y,z}$ are Twiss parameters in the y and z directions and satisfy $\gamma_{y,z} = (1 + \alpha_{y,z}^2)/\beta_{y,z}$. $E_0, \sigma_E, \sigma_x$ and $\epsilon_{y,z}$ are the nominal beam energy, the energy spread, the longitudinal beam length and transverse emittances. Twiss parameters and emittances are related with the transverse beam size $\sigma_{y,z}$ and the beam divergence $\theta_{y,z}$ by $\gamma_{y,z} = \theta_{y,z}^2/\epsilon_{y,z}$ and $\beta_{y,z} = \sigma_{y,z}^2/\epsilon_{y,z}$. We use initial electron beam parameters by $\sigma_{y,z}$, $\theta_{y,z}$, $\sigma_E$ and $E_0$ for the comparision convenience with experimental data.

Electrons are tracked with the synchrotron radiation code WAVE[28] developed at Helmholtz Zentrum Berlin, Germany. WAVE can calculate electrons trajectories and synchrotron radiation for arbitrary magnetic fields with high precision. Magnetic field inside a $12 \times 0.5 \times 0.5$ mm box with grid resolution ($\lambda_u/25$, $\lambda_u/50$, $\lambda_u/50$) in x, y, z direction, respectively, was loaded into WAVE for electron beam tracking and spectrum calculation. For the spectrum calculation, $5 \times 10^3$ electrons were considered.

## Acknowledgements

The authors acknowledge M.Scheer at HZB for fruitful discussions.This work was supported by the National Key R&D Program of China (2017YFA0403301), National Natural Science Foundation of China (11334013, 11721404, U1530150, 11805266), and the Key Program of CAS (XDB17030500, XDB16010200).

## Conflict of interests

The authors declare no competing financial interests.

## Contributions

J.H.T and L.M.C proposed the concept presented in this paper. J.H.T carried out the simulation and wrote the paper. D.Z.L. contributed to the magnetic field calculation. Y.F.L., B.J.Z., C.Q.Z., J.G.W. and D.Z.L. contributed to analysis of the results and write the paper. All authors discussed the results and commented on the paper.